# Tunable intraband optical conductivity and polarization-dependent epsilon-near-zero behavior in black phosphorus

Souvik Biswas[1], William S. Whitney[2], Meir Y. Grajower[1], Kenji Watanabe[3], Takashi Taniguchi[3], Hans A. Bechtel[4], George R. Rossman[5], Harry A. Atwater[1*]

1. Thomas J. Watson Laboratory of Applied Physics, California Institute of Technology, Pasadena, California 91125, United States
2. Department of Physics, California Institute of Technology, Pasadena, California 91125, United States
3. Research Center for Functional Materials, National Institute for Materials Science, 1-1 Namiki, Tsukuba 305-0044, Japan
4. Advanced Light Source Division, Lawrence Berkeley National Laboratory, Berkeley, California 94720, USA
5. Division of Geological and Planetary Sciences, California Institute of Technology, Pasadena, California 91125, United States

*Corresponding author: Harry A. Atwater (haa@caltech.edu)

**Abstract**

**Black phosphorus (BP) offers considerable promise for infrared and visible photonics. Efficient tuning of the bandgap and higher subbands in BP by modulation of the Fermi level or application of vertical electric fields has been previously demonstrated, allowing electrical control of its above bandgap optical properties. Here, we report modulation of the optical conductivity below the band-gap (5-15 μm) by tuning the charge density in a two-dimensional electron gas (2DEG) induced in BP, thereby modifying its free carrier dominated intraband response. With a moderate doping density of $7 \times 10^{12}$ cm$^{-2}$ we were able to observe a polarization dependent epsilon-near-zero behavior in the dielectric permittivity of BP. The intraband polarization sensitivity is intimately linked to the difference in effective fermionic masses along the two crystallographic directions, as confirmed by our measurements. Our results suggest the potential of multilayer BP to allow new optical functions for emerging photonics applications.**

# MAIN TEXT
## Introduction:

Hyperbolic photonic materials, in which the dielectric permittivities associated with different polarization directions have opposite signs, present a unique platform to engineer extremely strong anisotropic light-matter interactions and tailor novel topological properties of light (*1, 2*). They can enable a wide range of phenomena such as near field enhancement and modification of the local density of states of emitters(*3*), negative refraction(*4*), hyperlensing(*5*), super-Planckian thermal emission(*6*), sub diffraction light confinement(*7*), canalization of incident energy(*8*), and more. In addition to passive artificial metamaterial structures based on periodically arranged metal-dielectric stacks(*9*), hyperbolic dispersion has also been explored for a wide range of natural materials such as graphite, h-BN, WTe$_2$ in different spectral ranges (*10–12*). However, the idea of an electrically or optically tunable on-demand hyperbolic material is highly attractive for study of fundamental phenomena such as achieving active control of optical topological transitions, as well as applications in optical information processing and switching, and other functions. (*13*, *14*).

Two-dimensional electron gases (2DEG) in atomically thin materials with strong electro-optic susceptibility offer an ideal platform to achieve highly tunable light-matter interactions(*13*–*16*). These systems have established critical metrological standards in the field of condensed matter physics (such as the fine structure constant(*17*) and conductance quanta(*18*)) and have contributed to advances in photonics(*19*–*21*). Black phosphorus (BP), among other two-dimensional systems, has been heavily explored as an electronic platform for high mobility 2DEG(*22*–*27*), but very little is known about its optical properties. Bulk BP crystal possesses an anisotropic direct bandgap that is known to dramatically increase from 0.3eV to 2eV as the atomically thin limit (monolayer) is reached(*28*, *29*). In addition, the highly anisotropic band structure and optical properties of BP are extremely susceptible to perturbations in the local dielectric environment(*30*), temperature(*31*), electron/hole concentration in the 2DEG in BP(*32*), electric or magnetic field(*33*, *34*), strain(*35*, *36*), etc. While monolayer and few-layer BP can exhibit strong light-matter interactions by virtue of excitonic resonances in the visible-near infrared (IR)(*37*, *38*), multilayer BP holds more potential in the mid-IR because of its lower bandgap and stronger Drude weight(*39*). Quantum well electro-optic effects and its anisotropy near the band-edge have been studied recently in some detail in multilayer BP(*40*–*43*). However, absorption below the optical gap, which should be dominated by free carriers in the 2DEG, is still experimentally poorly understood. The free-carrier response of doped BP films can persist up to mid-infrared frequencies and can be approximated to first order by a Drude model(*44*, *45*). Similar behavior has been observed in graphene(*46*), but has not been explored in BP. Knowledge about the charge dynamics can provide us with an understanding of how quasiparticles in BP respond to infrared electromagnetic radiation, and the exact nature of their respective scattering and damping processes. A comprehensive understanding of the polarization-dependent, mid-IR optical properties of BP may facilitate the development of BP-based photonic devices, which hold promise for novel optoelectronic functions in emerging technology applications.

In this work, we report a comprehensive study of the optical conductivity of a 2DEG induced in multilayer BP for different hole and electron densities by performing reflection spectroscopy. Modulation of reflection was observed both above and below the BP band edge. While changes near or above the optical gap can be understood from an interplay of different electro-optic effects in BP (such as Pauli blocking, quantum confined Stark effect, etc.) and modelled using the Kubo formalism, modulation below the band gap is attributed primarily to changes in the intraband optical conductivity which has a Drude like frequency response ($\sigma = \frac{iD}{\pi(\omega + i\Gamma)}$). We measured the Drude weight (D) evolution and thus the change in the optical conductivity as a function of electron/hole concentration in the 2DEG. As predicted by theory, we observed anisotropy of the polarization resolved Drude response due to the difference in the effective mass of carriers along the two principal crystal axes. Changes in the intraband absorption imply a transfer of the spectral weight from the interband transitions, thereby preserving the oscillator f-sum rule for solid-state systems, $\int_0^\infty \Delta\sigma'(\omega)d\omega = 0$. Finally, from the extracted complex polarized dielectric function for BP below the band-edge, we were able to identify an epsilon near zero (ENZ) like regime along the armchair (AC) direction and hence, a transition from dielectric to metallic like response. No such transition was seen for the zigzag (ZZ) direction, confirming that doped multilayer BP is indeed an ideal system to host plasmons with tunable in-plane hyperbolic dispersion in the mid-IR. Furthermore, our results indicate a possible gate tunable optical topological transition for TM polarized light from hyperbolic to elliptical, which suggests interesting opportunities for multilayer BP in mid-IR photonic applications(*47*).

## Results

**Optical and electrical characterization of multilayer BP field effect heterostructure:**

We employed Fourier-transform infrared micro-spectroscopy to measure reflection spectra of multilayer BP devices. An typical field effect heterostructure constructed using van der Waals assembly technique is shown in Fig. 1(c), where the BP is 18.7 nm and the top and bottom h-BN are 36.8 nm and 36.4 nm respectively. The carrier density in BP was tuned by applying a gate voltage across the bottom hBN and $SiO_2$ (285nm). Such a geometry, schematically illustrated in the inset of Fig. 1(b), allowed independent electrical and optical characterization of the induced 2DEG in BP. Polarized Raman spectroscopy was used to identify the AC and ZZ axes of BP as indicated in the optical image of the device. All optical and electrical measurements were performed in ambient at room temperature.

The device was first characterized electrically by performing two terminal transconductance measurements as shown in Fig. 1(e)-(g). Ambipolar conduction was observed with an on/off ratio of ~100 (limited by contact resistance) and 17V was identified as the minimal conductance point (MCP), thereafter assumed to be the flat band condition in BP. A hole mobility of 1107 $cm^2$/V-s and an electron mobility of 412 $cm^2$/V-s were obtained at low doping levels, corresponding to scattering times on the order of ~5-10 meV. We then measured the reflection spectrum for the same device at the MCP with light polarized along the AC direction. This spectrum is normalized to that of optically thick Au (approx. 500 nm) evaporated on the same sample as a reference surface, the result of which is shown in Fig. 1(d). Three prominent features dominate the spectrum – a narrow hBN phonon around 1370 $cm^{-1}$, a broad dominant $SiO_2$ phonon around 1100 $cm^{-1}$ (recent studies(*13*) show multiple phonon contributions in $SiO_2$) and the beginning of band edge absorption around 3000 $cm^{-1}$ convoluted with an interference dip coming from the entire stack. A separate figure in the Supplementary Information shows the spectral features interband absorption for multiple subbands. As shown in Figs. 1(h) and 1(i), these reflection spectra can be heavily modified under positive or negative gate voltages.

Modelling the optical conductivity of the BP electron/hole gas allows us to gain an understanding of the quasiparticle dynamics under applied voltage. Our BP flakes are between 10-20 nm thick and described by a sheet conductivity $\sigma$ since the effective modulation is confined to only 2-3nm from the interface of BP/b-hBN. The thickness of this modulated region was estimated from the results of band bending calculations which are detailed in the Supplementary Information. This sheet conductivity has contributions from both interband and intraband processes, given as $\sigma = \sigma_{interband} + \sigma_{intraband} = \sigma_1(\omega) + i\sigma_2(\omega)$. The interband contribution accounts for absorption above the band-edge, including all subbands, while the intraband part accounts for free carrier response. One can explicitly calculate for optical conductivity using the Kubo formalism as follows(*45, 48*):

$$\sigma_{interband} = -i \frac{g_s \hbar e^2}{(2\pi)^2} \sum_{ss'jj'} \int dk \, \frac{f(E_{sjk}) - f(E_{s'j'k'})}{E_{sjk} - E_{s'j'k'}} * \frac{\langle \phi_{sjk} | \widehat{v_\alpha} | \phi_{s'j'k'} \rangle \langle \phi_{s'j'k'} | \widehat{v_\beta} | \phi_{sjk} \rangle}{E_{sjk} - E_{s'j'k'} + \hbar\omega + i\eta} \quad \text{(E1)}$$

$$\sigma_{intraband,j} = \frac{iD_j}{\pi(\omega + \frac{i\eta}{\hbar})}, \, D_j = \pi e^2 \sum_{i=1}^{N} \frac{n_i}{m^*_{i,j}} \quad \text{(E2)}$$

Here, $\widehat{v_{\alpha,\beta}}$ is the velocity operator defined as $\hbar^{-1} \partial_{k_{\alpha,\beta}} H$, $g_s = 2$ is used to denote the spin degeneracy, f(E) is the Fermi-Dirac distribution function; the indices s(s′) refer to conduction (valence) bands and the indices j(j′) refer to the subbands. H is the low-energy in-plane Hamiltonian around the Γ point, $E_{sjk}$, $\phi_{sjk}$ are the eigen-energies and eigenfunctions of H; $m^*_{i,j}$ represents the effective mass of carriers in each subband (i) along a specific crystal orientation j, $n_i$ represents the change density in each subband and $\eta$ is a phenomenological damping term.

Electrostatic doping of BP primarily brings about two fundamental changes in the optical response: the emergence of a strong intraband component in the mid to far infrared and a shift of the optical gap (interband transitions), shown schematically in Fig. 1(j). A combination of multiple electro-optical effects at the band-edge have been shown to explain the observed modulation (more details are available in the Supplementary Information). All of the observed reflection modulation spectra exhibit strong anisotropy with respect to the BP crystal axes under AC and ZZ polarized illumination. This strong anisotropy is predicted by theory, and results from the puckered honeycomb lattice crystal structure of phosphorene(*49*). Our results in Figs. 1(h) and 1(i) indicate significant optical modulation in the 2DEG and are in excellent agreement with results from previous studies(*40-43*).

**Low energy doping dependent intraband response in multilayer BP:**

We now turn our attention to the low photon energy regime, which is dominated by the intraband conductivity of BP. Fig. 2 describes this response, the understanding of which is a central result of this paper. Figs. 2(a) and (b) show reflectance spectra (normalized as before) for light polarized along the AC and ZZ axes, respectively. Both electron and hole doping can modify the free carrier response of the 2DEG. As doping increases, a strong spectral feature is observed to appear around the characteristic hBN (~1360 cm$^{-1}$) and SiO$_2$ (~1100 cm$^{-1}$) intrinsic phonon peaks with both electron and hole doping. We propose that this feature results from an increase in the free carrier density, which increases the Drude conductivity and thus modifies the optical properties of BP. This broad intraband modulation interferes with the previously described hBN/SiO$_2$ phonons, giving rise to an absorption line shape with a Fano-like modulation in the hBN/SiO$_2$ phonon regime. We hypothesize that this asymmetric Fano-like resonance shape(*50, 51*) indicates optical coupling between the narrow phonon resonances and the weak free carrier absorption continuum. To better understand the nature of the line shape, we performed thin film transfer matrix calculations to fit the spectra and account for multiple reflections and interferences in the heterostructure stack. Our model incorporates a Drude-like function for the intraband optical conductivity of the BP 2DEG, given by equation (E2), with which we are able to extract the Drude weight as a function of doping. Assuming a simple parallel plate capacitor model, we can estimate the doping density at each gate voltage. For undoped BP we assume a charge density of $10^{11}$/cm$^2$ to account for the finite MCP response (coming from any defects or trapped charges). The contribution to the linewidth of the imaginary component of Drude conductivity from dephasing associated with finite scattering times was assumed to be on the order of that obtained from DC transport measurements (approx. 5 meV) which is a valid approximation in the energy ranges considered here. The possible sources of scattering include electron-phonon coupling, electron-electron repulsion and interaction with defects and impurities. Studies have shown that crystals of layered materials on substrates with strong phonons can also show losses from electron-surface polar phonon coupling(*52*). There have also been reports of DC transport mobilities which are not well correlated with optical scattering times, or which even show an anti-correlation(*53*). Further fundamental spectroscopic studies at far-IR (THz) frequencies will be required to further elucidate these low energy scattering mechanisms in BP as a function of doping and temperature. Figs. 2(c) and (d) summarize the fitted results without any offset to better understand the impact of doping on the lineshape of reflectance modulation. Excellent agreement between experimental data and transfer matrix simulations is visible in Figs. 2(a) and (b), which indicates that the intraband (Drude) model suffices to explain the reflectance modulation observed at photon energies well below the band-edge.

Figs. 2(e) and 2(f) also show in false colors the changes in the reflection modulation for AC and ZZ polarization as a function of electron/hole doping density assuming a constant effective mass for BP. Reflection/transmission in highly sub-wavelength BP films are mostly dominated by the losses in the material and thus, it is important to note that we do not incorporate the interband region in our Drude modelling of the sub-bandgap response because we are working much below the (even the Stark shifted ) band-gap, where the influence of interband losses is almost negligible to first order. Similar assumptions have been experimentally validated for studies on graphene(*46, 54*). Also, it should be noted that while the interband anisotropy is primarily governed by the parity of wavefunctions and subsequent selection rules coming from dipole matrix elements in BP, intraband anisotropy stems from the difference in fermionic effective mass along the two crystallographic axes.

**Measurement of the multilayer BP complex permittivity and tunable epsilon-near-zero and hyperbolicity:**

Figures 3(a)-(d) illustrate the experimental real and imaginary parts of the dielectric function (obtained as $\epsilon_{AC/ZZ}(\omega) = \epsilon_{\infty(AC/ZZ)} + \frac{i\sigma_{AC/ZZ}(\omega)}{t\epsilon_0 \omega}$, where t = thickness of the 2DEG (2.9nm), $\epsilon_{\infty(AC/ZZ)}$ accounts for oscillators not captured in our "Drude" spectral window) for BP at different doping densities under polarized excitation conditions along the AC and ZZ directions. At higher energies the dielectric function is dominated by subband transitions whose oscillator strength diminishes upon doping primarily due to Pauli blocking, along with the aforementioned electro-optic effects. The lower energy response is mostly dominated by free carriers. We observe a strong modulation of the dielectric function below the band gap with doping density, indicating that free carrier response is a significant effect in the mid-IR range. An important finding of our study is the appearance of an epsilon near zero region (ENZ) in BP for higher gate voltages / charge densities along the AC direction, where the real part of the permittivity transitions from positive to negative. A false color plot showing the variation of the modelled real part of the dielectric permittivity with doping density along the AC direction is shown in Fig. 3(e). The ENZ region is seen to systematically shift to higher photon energies with increased doping. Here the effective mass of BP is assumed to be $0.14m_0$, independent of doping density, for the sake of simplicity. No such negative permittivity region was identified for the ZZ direction measurements, implying extreme bianisotropy and the possibility to generate surface plasmon modes and in-plane hyperbolic photonic dispersion in BP. We further calculate the isofrequency contours for in-plane plasmons (TM polarized surface modes)*(10)* for two different doping densities (one on the electron side and one on the hole side) at 750 cm$^{-1}$ to show that the hyperbolic dispersion is electrically tunable. Electrically tunable hyperbolic dispersion, illustrated in Fig. 3(f) has intriguing implications, suggesting opportunities for active hyperbolic plasmonics and photonics. We note that the doping density achieved here is modest (~$7 \times 10^{12}$ cm$^{-2}$), and higher doping densities with larger κ dielectrics may enable the hyperbolic dispersion regime to move to shorter wavelengths. In the frequency regime accessible in our measurements, we do not observe a negative real permittivity along the ZZ direction and we expect it to occur at much lower frequencies (<500 cm$^{-1}$), which indicates that any surface plasmon modes at frequencies above 500 cm$^{-1}$ will inherit a hyperbolic dispersion (as shown in Fig. 3(f)) whereas those below will inherit an elliptical dispersion thereby undergoing an topological transition in photonic dispersion. It should be possible to electrically tune the transition point, as indicated by our results. Theoretical studies of surface plasmons in BP and their corresponding dispersion relations have been discussed elsewhere(*8*); however our results provide a concrete step in that direction.

**Determination of carrier effective masses in a multilayer BP 2DEG:**

Finally, we use our experimental results to obtain carrier effective masses that can be compared with results from theory as shown in Fig. 4(a) and (b). We see qualitatively good agreement with theory ($m_{eff} \approx 0.14m_0$) for our results along the AC axis, in fact our extracted fermionic effective mass is slightly heavier than the previously theoretically calculated results. We speculate this could be an interplay of two effects. Firstly, in our BP thin films the 2DEG is highly confined and thus the band dispersion is modified, leading to heavier confined fermions(*26*). Additionally, when the Fermi level moves into either the conduction or valence band with gating, we access not only the minima and maxima of the first subbands in the conduction and valence band respectively, but also the higher subbands (because of the broad Fermi-Dirac tail at 300K). In BP for higher lying subbands along the AC axis, the effective mass increases gradually, as given by $m_{AC}^i = \frac{\hbar^2}{\frac{2\gamma^2}{\delta^i}+\eta}$, where $\delta^i$ is the subband transition energy, $\gamma$ denotes the effective coupling between the conduction/valence bands and $\eta$ is related to the in-plane dispersion of the bands(*44, 45*). It is possible that the carriers participating in the intraband transitions come from a mixture of the different subbands thus leading to an overall lower perceived effective mass. For the ZZ axis ($m_{eff} \approx 0.71m_0$), we see a slightly larger variation in the extracted effective mass between the electron and hole side. Electronic confinement leads to heavier fermions along the ZZ direction, which is in accordance with our optical measurements for hole conductivity. It is possible that non parabolic band effects give rise to a slightly lower effective mass for electrons, but further detailed analysis is needed to resolve the observed electron-hole asymmetry. It should be noted that for the ZZ direction the effective mass is not expected to depend on the subband index. As expected from theory, the ZZ carriers are found to be much heavier than carriers associated with transport along the AC direction. In order to verify all of our results, we performed similar measurements with an unpolarized beam on the same sample (presented in Supplementary Information). We see highly consistent and reproducible interband conductivity modulation effects for light polarized along the AC or ZZ axis. Below the band-edge we see a systematic strengthening of the Fano-like response for both electron and hole doping in the wavelength range near the hBN and $SiO_2$ phonons. Fits to those data sets are in excellent agreement with our theoretical analysis. Additionally, we find that the effective mass obtained is approximately the average of the in-plane effective masses along the AC and ZZ directions. This indicates that carriers along both the axes participate in the intraband response and is consistent with BP transport measurements. Subtle deviations are expected to occur in the estimation of the effective mass because of the non-symmetric nature of the unpolarized beam in our setup (it is slightly elliptically polarized) and possibly because of complex cross-scattering mechanisms between the two crystallographic directions. Overall, both polarized and unpolarized measurements are in good agreement with theory. We note that it is difficult to eliminate small conductivity changes due to local fluctuations in the charge density arising from charge puddles (bubbles/grain boundaries/defects) in heterostructures since our measurements probe a large area (approx. 2500 μm²)(*55*). Additionally, studies have shown some systems to have an energy dependent quasiparticle scattering rate owing to strong electron-electron and electron-phonon interactions(*56, 57*). Such interactions can cause additional broadening in both the interband and the intraband absorption features and have not been considered in our data analysis. We also note that in thick BP films (> 5 nm) like those measured here, excitonic effects are negligible and hence have not been taken into account. We performed measurements on three other BP heterostructures and saw very similar behavior.

**Conclusions:**

We have experimentally explored the below bandgap optical response in gated multilayer BP heterostructures and identified the dominant contribution to be a Drude-like optical conductivity due to free carriers in the 2DEG. We find that interband transitions play a negligible role in the low photon energy response for BP, which greatly simplifies modelling of the optical conductivity of BP 2DEGs and subsequent photonic devices. We have measured the anisotropy in the intraband optical conductivity of BP by performing polarized reflection measurements and extracted the effective masses along the two crystallographic axes as a function of charge density. Our intraband optical conductivity results are consistent with any changes in the interband regime and DC transport measurements. Moreover, we demonstrated the existence of a plasmonic regime with electrically tunable hyperbolic dispersion and an ENZ regime. We also identified the wavelength regime for the onset of a topological transition for BP surface plasmons between hyperbolic and elliptical dispersion. Our results provide a foundation for a range of future research directions investigating BP as a strongly bianisotropic (or hyperbolic) mid-IR material for applications such as plasmonics, molecular fingerprinting, sensing and tailoring thermal emission. They also pose important questions about the different scattering mechanisms in BP amidst a complex phase space of doping, thickness, substrate and temperature, and the nature of the mid-IR to THz response in BP, which motivates future work. Finally, our demonstration of BP as a naturally occurring material with tunable hyperbolic dispersion and bianisotropy suggests applications in novel photonics such as active polarization-sensitive infrared metasurfaces.

## Materials and Methods

### Device fabrication -

Black phosphorus (crystals from hqgraphene) was exfoliated using blue Nitto tape onto polydimethylsiloxane (PDMS) stamps and suitable flakes were identified. Flakes were chosen based on the following parameters- contrast in the false-color microscope images in transmission (lower contrast flakes were thinner), size (>30μm x 30μm for spectroscopic measurements) and shape (rectangular/square shapes with easily identifiable axes were chosen for ease of fabrication and polarization measurements). Thin hBN flakes (bulk crystals from NIMS) were exfoliated onto $SiO_2$/Si (double side polished) and identified on the basis of their colors. The thickness of each flake was later checked with atomic force microscopy (AFM). Large flakes of hBN were selected so that the BP would be fully encapsulated (this is important to prevent any ambient degradation of BP). Once the bottom hBN was identified, an appropriate BP flake was transferred from PDMS on to hBN using the dry transfer technique at 60℃ slowly ensuring a clean interface; the cleanliness of this interface is the most important since it hosts the 2DEG in BP. Another large BN flake was then picked up with (poly-carbonate) PC/PDMS at 90℃ and then dropped on the BP/hBN stack at 180℃ slowly ensuring that two edges of BP were exposed. For the device discussed in the main manuscript, the exposed edges were along the AC direction for maximal functionality and accurate estimation of mobilities along the same. The PC was then washed off with chloroform for 10 minutes and IPA for 30s and dried with $N_2$. PMMA 950A4 was spun at 3000r.p.m. for 1 minute (thickness ~200nm) and then baked at 180ºC for 3 minutes for electron-beam lithography. Electrical contacts to the BP were written with electron beam lithography at an acceleration voltage of 100 keV and a beam current of 1nA (area dosage of 1350μC/cm$^2$), developed in MIBK:IPA 1:3 for 1 minute and then rinsed in IPA for 30s (followed by blow dry with $N_2$). Metal (Ti/Au 5nm/95nm) was evaporated at 0.5Å/s and 1Å/s respectively, at a base pressure of 7e-8 Torr, ensuring electrical contact. All the processing steps were done inside the glovebox to prevent degradation, except the e-beam lithography and the step between developing the resist and metal evaporation. The time between developing and loading in the loadlock for metal evaporation was less than 2 minutes.

**Measurements-**

Once the device fabrication was complete, the chip was wire bonded onto a custom made printed circuit board and mounted on a polarization resolved Raman spectroscopy setup. Polarized Raman measurements were used to confirm the axes of BP. Two Keithley 2400s were used (automated with LabView scripts) for the DC transport measurements, which involved two different methods of measurement: 1. At a constant source-drain voltage the gate voltage was swept from 0 upto the maximum positive voltage, back to 0, then to the negative maximum voltage and then to 0 at 0.1V/s, 2. At a constant gate voltage, the source-drain voltage was swept in a similar fashion. Both modes of measurements show highly similar MCP and on/off ratios and linear source-drain transport showing Ohmic contacts (at room temperature). Then the PCB was mounted onto an (Nicolet iS50) FTIR-coupled with an infrared microscope setup. A 15x Cassegrain objective with NA=0.58 was used for all the measurements. A KBr beam-splitter was used for the interferometry and a KBr wire-grid linear polarizer for polarization resolved reflection measurements. Reflection spectra were acquired using a liquid nitrogen cooled HgCdTe detector from 600-8000 cm$^{-1}$ with a resolution of 4 cm$^{-1}$. The linear polarizer was used in the optical path before the sample to polarize the incident light along the two principal axes of BP. Voltage was applied with a Keithley 2400. During the optical measurements the source and drain were held at the same voltage and the back gate voltage was systematically varied from -90V to +90V (higher voltages were avoided to prevent damage to the device, but based on our previous experience most of these devices may survive upto much higher voltages since breakdown usually occurs above 120-140V). The BP thickness was examined after all the measurements were done since we don't have an AFM in the glovebox. Good correlation was found between the optical contrast and AFM height checks. Also, the hBN top and bottom flakes were inspected with infrared spectroscopy before the assembly of the device to extract the phonon parameters.

It should also be noted that for such long wavelength infrared measurements, systematic drifts in the system can lead to artificial features and in order to eliminate any such effect we adopted the following techniques – 1. After taking data at a specific voltage, the gate bias was taken to the MCP level and a spectrum was taken immediately which served as a baseline/reference for that specific voltage; 2. The manner in which voltages were scanned were completely randomized to ensure the modulation seen is not a time-dependent drift in the system; 3. The whole process was repeated on the same device 4-5 times to confirm a systematic change was seen.

**Supplementary Materials:**
Fig. S1. Differential reflectance change in BP.
Fig. S2. Polarized Raman spectra measured for the BP device presented in the main manuscript.
Fig. S3. Schematic of different electro-optic effects occurring at energies near and above the band-edge of a multilayer BP thin film.
Fig. S4. Unpolarized reflection modulation measurements and calculations.
Tab. T1. Bounds on effective mass in BP.
Fig. S5. Calculation of transition energies in 18.68nm BP and corresponding effective mass along the Armchair direction.
Fig. S6. Refractive index data adopted for SiO$_2$.
Fig. S7. Fano like response for BP/SiO$_2$ system and BP/hBN system.
Fig. S8. Extracted refractive index of BP 2DEG (t=2.9 nm) as a function of voltage and polarization.
Fig. S9. AFM line scans for top hBN, BP flake and bottom hBN.
Fig. S10. Charge density induced in BP as calculated from parallel plate capacitor model.
Fig. S11. Thomas Fermi screening calculation in BP.

References:

1. A. Poddubny, I. Iorsh, P. Belov, Y. Kivshar, Hyperbolic metamaterials. *Nature Photonics*. **7** (2013), pp. 958–967.

2. H. N. S. Krishnamoorthy, Z. Jacob, E. Narimanov, I. Kretzschmar, V. M. Menon, Topological Transitions in Metamaterials. *Science*. **336**, 205–209 (2012).

3. D. Lu, J. J. Kan, E. E. Fullerton, Z. Liu, Enhancing spontaneous emission rates of molecules using nanopatterned multilayer hyperbolic metamaterials. *Nature Nanotechnology*. **9**, 48–53 (2014).

4. A. A. High, R. C. Devlin, A. Dibos, M. Polking, D. S. Wild, J. Perczel, N. P. de Leon, M. D. Lukin, H. Park, Visible-frequency hyperbolic metasurface. *Nature*. **522**, 192–196 (2015).

5. Z. Liu, H. Lee, Y. Xiong, C. Sun, X. Zhang, Far-field optical hyperlens magnifying sub-diffraction-limited objects. *Science*. **315**, 1686 (2007).

6. S. A. Biehs, M. Tschikin, R. Messina, P. Ben-Abdallah, Super-Planckian near-field thermal emission with phonon-polaritonic hyperbolic metamaterials. *Applied Physics Letters*. **102**, 131106 (2013).

7. J. D. Caldwell, A. v. Kretinin, Y. Chen, V. Giannini, M. M. Fogler, Y. Francescato, C. T. Ellis, J. G. Tischler, C. R. Woods, A. J. Giles, M. Hong, K. Watanabe, T. Taniguchi, S. A. Maier, K. S. Novoselov, Sub-diffractional volume-confined polaritons in the natural hyperbolic material hexagonal boron nitride. *Nature Communications*. **5**, 5221 (2014).

8. D. Correas-Serrano, J. S. Gomez-Diaz, A. A. Melcon, A. Alù, Black phosphorus plasmonics: anisotropic elliptical propagation and nonlocality-induced canalization. *Journal of Optics*. **18**, 104006 (2016).

9. T. Xu, H. J. Lezec, Visible-frequency asymmetric transmission devices incorporating a hyperbolic metamaterial. *Nature Communications*. **5**, 1–7 (2014).

10. C. Wang, S. Huang, Q. Xing, Y. Xie, C. Song, F. Wang, H. Yan, Van der Waals thin films of WTe2 for natural hyperbolic plasmonic surfaces. *Nature Communications*. **11**, 1–9 (2020).

11. S. Dai, Z. Fei, Q. Ma, A. S. Rodin, M. Wagner, A. S. McLeod, M. K. Liu, W. Gannett, W. Regan, K. Watanabe, T. Taniguchi, M. Thiemens, G. Dominguez, A. H. Castro Neto, A. Zettl, F. Keilmann, P. Jarillo-Herrero, M. M. Fogler, D. N. Basov, Tunable phonon polaritons in atomically thin van der Waals crystals of boron nitride. *Science*. **343**, 1125–1129 (2014).

12. J. Sun, J. Zhou, B. Li, F. Kang, Indefinite permittivity and negative refraction in natural material: Graphite. *Applied Physics Letters*. **98**, 101901 (2011).

13. G. X. Ni, A. S. McLeod, Z. Sun, L. Wang, L. Xiong, K. W. Post, S. S. Sunku, B. Y. Jiang, J. Hone, C. R. Dean, M. M. Fogler, D. N. Basov, Fundamental limits to graphene plasmonics. *Nature*. **557**, 530–533 (2018).

14. L. Ju, B. Geng, J. Horng, C. Girit, M. Martin, Z. Hao, H. A. Bechtel, X. Liang, A. Zettl, Y. R. Shen, F. Wang, Graphene plasmonics for tunable terahertz metamaterials. *Nature Nanotechnology*. **6**, 630–634 (2011).

15. Z. Fei, A. S. Rodin, G. O. Andreev, W. Bao, A. S. McLeod, M. Wagner, L. M. Zhang, Z. Zhao, M. Thiemens, G. Dominguez, M. M. Fogler, A. H. Castro Neto, C. N. Lau, F. Keilmann, D. N. Basov, Gate-tuning of graphene plasmons revealed by infrared nano-imaging. *Nature*. **486**, 82–85 (2012).

16. J. Chen, M. Badioli, P. Alonso-González, S. Thongrattanasiri, F. Huth, J. Osmond, M. Spasenović, A. Centeno, A. Pesquera, P. Godignon, A. Zurutuza Elorza, N. Camara, F. J. García, R. Hillenbrand, F. H. L. Koppens, Optical nano-imaging of gate-tunable graphene plasmons. *Nature*. **487**, 77–81 (2012).

17. K. v. Klitzing, G. Dorda, M. Pepper, New Method for High-Accuracy Determination of the Fine-Structure Constant Based on Quantized Hall Resistance. *Physical Review Letters*. **45**, 494–497 (1980).

18. B. Jeckelmann, B. Jeanneret, The quantum Hall effect as an electrical resistance standard. *Reports on Progress in Physics*. **64**, 1603 (2001).

19. M. Liu, X. Yin, E. Ulin-Avila, B. Geng, T. Zentgraf, L. Ju, F. Wang, X. Zhang, A graphene-based broadband optical modulator. *Nature*. **474**, 64–67 (2011).

20. M. Liu, X. Yin, X. Zhang, Double-layer graphene optical modulator. *Nano Letters*. **12**, 1482–1485 (2012).

21. B. Sensale-Rodriguez, R. Yan, M. M. Kelly, T. Fang, K. Tahy, W. S. Hwang, D. Jena, L. Liu, H. G. Xing, Broadband graphene terahertz modulators enabled by intraband transitions. *Nature Communications*. **3**, 780 (2012).

22. G. Long, D. Maryenko, J. Shen, S. Xu, J. Hou, Z. Wu, W. K. Wong, T. Han, J. Lin, Y. Cai, R. Lortz, N. Wang, Achieving Ultrahigh Carrier Mobility in Two-Dimensional Hole Gas of Black Phosphorus. *Nano Letters*. **16**, 7768–7773 (2016).

23. N. Gillgren, D. Wickramaratne, Y. Shi, T. Espiritu, J. Yang, J. Hu, J. Wei, X. Liu, Z. Mao, K. Watanabe, T. Taniguchi, M. Bockrath, Y. Barlas, R. K. Lake, C. N. Lau, Gate tunable quantum oscillations in air-stable and high mobility few-layer phosphorene heterostructures. *2D Materials*. **2**, 011001 (2015).

24. X. Chen, Y. Wu, Z. Wu, Y. Han, S. Xu, L. Wang, W. Ye, T. Han, Y. He, Y. Cai, N. Wang, High-quality sandwiched black phosphorus heterostructure and its quantum oscillations. *Nature Communications*. **6**, 7315 (2015).

25. L. Li, Y. Yu, G. J. Ye, Q. Ge, X. Ou, H. Wu, D. Feng, X. H. Chen, Y. Zhang, Black phosphorus field-effect transistors. *Nature Nanotechnology*. **9**, 372–377 (2014).

26. L. Li, G. J. Ye, V. Tran, R. Fei, G. Chen, H. Wang, J. Wang, K. Watanabe, T. Taniguchi, L. Yang, X. H. Chen, Y. Zhang, Quantum oscillations in a two-dimensional electron gas in black phosphorus thin films. *Nature Nanotechnology*. **10**, 608-613 (2015).

27. L. Li, F. Yang, G. J. Ye, Z. Zhang, Z. Zhu, W. Lou, X. Zhou, L. Li, K. Watanabe, T. Taniguchi, K. Chang, Y. Wang, X. H. Chen, Y. Zhang, Quantum Hall effect in black phosphorus two-dimensional electron system. *Nature Nanotechnology*. **11**, 593-597 (2016).

28. A. Castellanos-Gomez, Isolation and characterization of few-layer black phosphorus. *2D Mater*. **1**, 025001 (2014).

29. V. Tran, R. Soklaski, Y. Liang, L. Yang, Layer-controlled band gap and anisotropic excitons in few-layer black phosphorus. *Physical Review B*. **89**, 235319 (2014).

30. D. Y. Qiu, F. H. da Jornada, S. G. Louie, Environmental Screening Effects in 2D Materials: Renormalization of the Bandgap, Electronic Structure, and Optical Spectra of Few-Layer Black Phosphorus. *Nano Letters*. **17**, 4706–4712 (2017).

31. G. Zhang, S. Huang, F. Wang, Q. Xing, C. Song, C. Wang, Y. Lei, M. Huang, H. Yan, The optical conductivity of few-layer black phosphorus by infrared spectroscopy. *Nature communications*. **11**, 1847 (2020).

32. R. Xu, S. Zhang, F. Wang, J. Yang, Z. Wang, J. Pei, Y. W. Myint, B. Xing, Z. Yu, L. Fu, Q. Qin, Y. Lu, Extraordinarily Bound Quasi-One-Dimensional Trions in Two-Dimensional Phosphorene Atomic Semiconductors. *ACS Nano*. **10**, 2046–2053 (2016).

33. A. Chaves, T. Low, P. Avouris, D. Çaklr, F. M. Peeters, Anisotropic exciton Stark shift in black phosphorus. *Physical Review B*. **91**, 155311 (2015).

34. X. Zhou, W. K. Lou, F. Zhai, K. Chang, Anomalous magneto-optical response of black phosphorus thin films. *Physical Review B*. **92**, 165405 (2015).

35. S. Huang, G. Zhang, F. Fan, C. Song, F. Wang, Q. Xing, C. Wang, H. Wu, H. Yan, Strain-tunable van der Waals interactions in few-layer black phosphorus. *Nature Communications*. **10**, 1–7 (2019).

36. J. Quereda, P. San-Jose, V. Parente, L. Vaquero-Garzon, A. J. Molina-Mendoza, N. Agraït, G. Rubio-Bollinger, F. Guinea, R. Roldán, A. Castellanos-Gomez, Strong Modulation of Optical Properties in Black Phosphorus through Strain-Engineered Rippling. *Nano Letters*. **16**, 2931–2937 (2016).

37. G. Zhang, A. Chaves, S. Huang, F. Wang, Q. Xing, T. Low, H. Yan, Determination of layer-dependent exciton binding energies in few-layer black phosphorus. *Science Advances*. **4**, eaap9977 (2018).

38. L. Li, J. Kim, C. Jin, G. J. Ye, D. Y. Qiu, F. H. da Jornada, Z. Shi, L. Chen, Z. Zhang, F. Yang, K. Watanabe, T. Taniguchi, W. Ren, S. G. Louie, X. H. Chen, Y. Zhang, F. Wang, Direct observation of the layer-dependent electronic structure in phosphorene. *Nature Nanotechnology*. **12**, 21–25 (2017).

39. C. Lin, R. Grassi, T. Low, A. S. Helmy, Multilayer Black Phosphorus as a Versatile Mid-Infrared Electro-optic Material. *Nano Letters*. **16**, 1683–1689 (2016).

40. W. S. Whitney, M. C. Sherrott, D. Jariwala, W. H. Lin, H. A. Bechtel, G. R. Rossman, H. A. Atwater, Field Effect Optoelectronic Modulation of Quantum-Confined Carriers in Black Phosphorus. *Nano Letters*. **17**, 78–84 (2020).

41. M. C. Sherrott, W. S. Whitney, D. Jariwala, S. Biswas, C. M. Went, J. Wong, G. R. Rossman, H. A. Atwater, Anisotropic Quantum Well Electro-Optics in Few-Layer Black Phosphorus. *Nano Letters*. **19**, 269–276 (2019).

42. C. Chen, X. Lu, B. Deng, X. Chen, Q. Guo, C. Li, C. Ma, S. Yuan, E. Sung, K. Watanabe, T. Taniguchi, L. Yang, F. Xia, Widely tunable mid-infrared light emission in thin-film black phosphorus. *Science Advances*. **6**, eaay6134 (2020).

43. R. Peng, K. Khaliji, N. Youngblood, R. Grassi, T. Low, M. Li, Midinfrared Electro-optic Modulation in Few-Layer Black Phosphorus. *Nano Letters*. **17**, 6315–6320 (2017).

44. T. Low, A. S. Rodin, A. Carvalho, Y. Jiang, H. Wang, F. Xia, A. H. Castro Neto, Tunable optical properties of multilayer black phosphorus thin films. *Physical Review B*. **90**, 075434 (2014).

45. T. Low, R. Roldán, H. Wang, F. Xia, P. Avouris, L. M. Moreno, F. Guinea, Plasmons and Screening in Monolayer and Multilayer Black Phosphorus. *Physical Review Letters*. **113**, 106802 (2014).

46. Y. C. Chang, C. H. Liu, C. H. Liu, S. Zhang, S. R. Marder, E. E. Narimanov, Z. Zhong, T. B. Norris, Realization of mid-infrared graphene hyperbolic metamaterials. *Nature Communications*. **7**, 10568 (2016).

47. J. S. Gomez-Diaz, M. Tymchenko, A. Alù, Hyperbolic Plasmons and Topological Transitions over Uniaxial Metasurfaces. *Physical Review Letters*. **114**, 233901 (2015).

48. T. Low, A. S. Rodin, A. Carvalho, Y. Jiang, H. Wang, F. Xia, A. H. Castro Neto, Tunable optical properties of multilayer black phosphorus thin films. *Physical Review B. **90**, 075434 (2014)*.

49. H. Asahina, A. Morita, Band structure and optical properties of black phosphorus. *Journal of Physics C: Solid State Physics.* **17**, 11 (1984).

50. P. Fan, Z. Yu, S. Fan, M. L. Brongersma, Optical Fano resonance of an individual semiconductor nanostructure. *Nature Materials*. **13**, 471–475 (2014).

51. B. Luk'Yanchuk, N. I. Zheludev, S. A. Maier, N. J. Halas, P. Nordlander, H. Giessen, C. T. Chong, The Fano resonance in plasmonic nanostructures and metamaterials. *Nature Materials*. **9**, 707–715 (2010).

52. B. Scharf, V. Perebeinos, J. Fabian, P. Avouris, Effects of optical and surface polar phonons on the optical conductivity of doped graphene. *Physical Review B.* **87**, 035414 (2013).

53. A. Principi, M. Carrega, M. B. Lundeberg, A. Woessner, F. H. L. Koppens, G. Vignale, M. Polini, Plasmon losses due to electron-phonon scattering: The case of graphene encapsulated in hexagonal boron nitride. *Physical Review B*. **90**, 165408 (2014).

54. Z. Fei, G. O. Andreev, W. Bao, L. M. Zhang, A. S. McLeod, C. Wang, M. K. Stewart, Z. Zhao, G. Dominguez, M. Thiemens, M. M. Fogler, M. J. Tauber, A. H. Castro-Neto, C. N. Lau, F. Keilmann, D. N. Basov, Infrared nanoscopy of dirac plasmons at the graphene-SiO2 interface. *Nano Letters*. **11**, 4701–4705 (2011).

55. A. Raja, L. Waldecker, J. Zipfel, Y. Cho, S. Brem, J. D. Ziegler, M. Kulig, T. Taniguchi, K. Watanabe, E. Malic, T. F. Heinz, T. C. Berkelbach, A. Chernikov, Dielectric disorder in two-dimensional materials. *Nature Nanotechnology*. **14**, 832–837 (2019).

56. J. González, F. Guinea, M. A. H. Vozmediano, Unconventional Quasiparticle Lifetime in Graphite. *Physical Review Letters*. **77**, 3589–3592 (1996).

57. Y. Liu, P. P. Ruden, Temperature-dependent anisotropic charge-carrier mobility limited by ionized impurity scattering in thin-layer black phosphorus. *Physical Review B*. **95**, 165446 (2017).

## Acknowledgments

**General**: The authors would like to thank Dr. Qiushi Guo regarding preliminary discussions about this project, Joeson Wong for discussions regarding transfer matrices, Ghazaleh Kafaie Shirmanesh for help with printed circuit boards for electrical measurements and Dr. Laura Kim for providing LabVIEW script to perform electrical measurements. **Funding:** The authors gratefully acknowledge support from the Department of Energy-Office of Science under grant DE-FG02-07ER46405. This research used resources of the Advanced Light Source, a U.S. DOE Office of Science User Facility under contract no. DE-AC02-05CH11231. K.W. and T.T. acknowledge support from the Elemental Strategy Initiative conducted by the MEXT, Japan, Grant Number JPMXP0112101001, JSPS KAKENHI Grant Numbers JP20H00354 and the CREST(JPMJCR15F3), JST. **Author contributions:** S.B., W.S.W. and H.A.A conceived the project. S.B. and W.S.W. worked on fabrication, measurements and analysis of preliminary data. S.B. fabricated, measured and analyzed data from final samples. M.Y.G. assisted in electrical and optical measurements. S.B., W.S.W. and M.Y.G. discussed the implications of the results. H.A.A. supervised the project. H.A.B. assisted with additional and complementary measurements done at the ALS, Berkeley. K.W. and T.T. provided hBN and BP crystals. G.R.R supervised some of the optical measurements. S.B. wrote the manuscript and all authors provided important feedback. **Competing interests:** The authors declare that they have no competing interests. **Data and materials availability:** All data required to evaluate the conclusions in the manuscript are available in the main text and/or Supplementary Materials. Additional data may be requested from the authors.

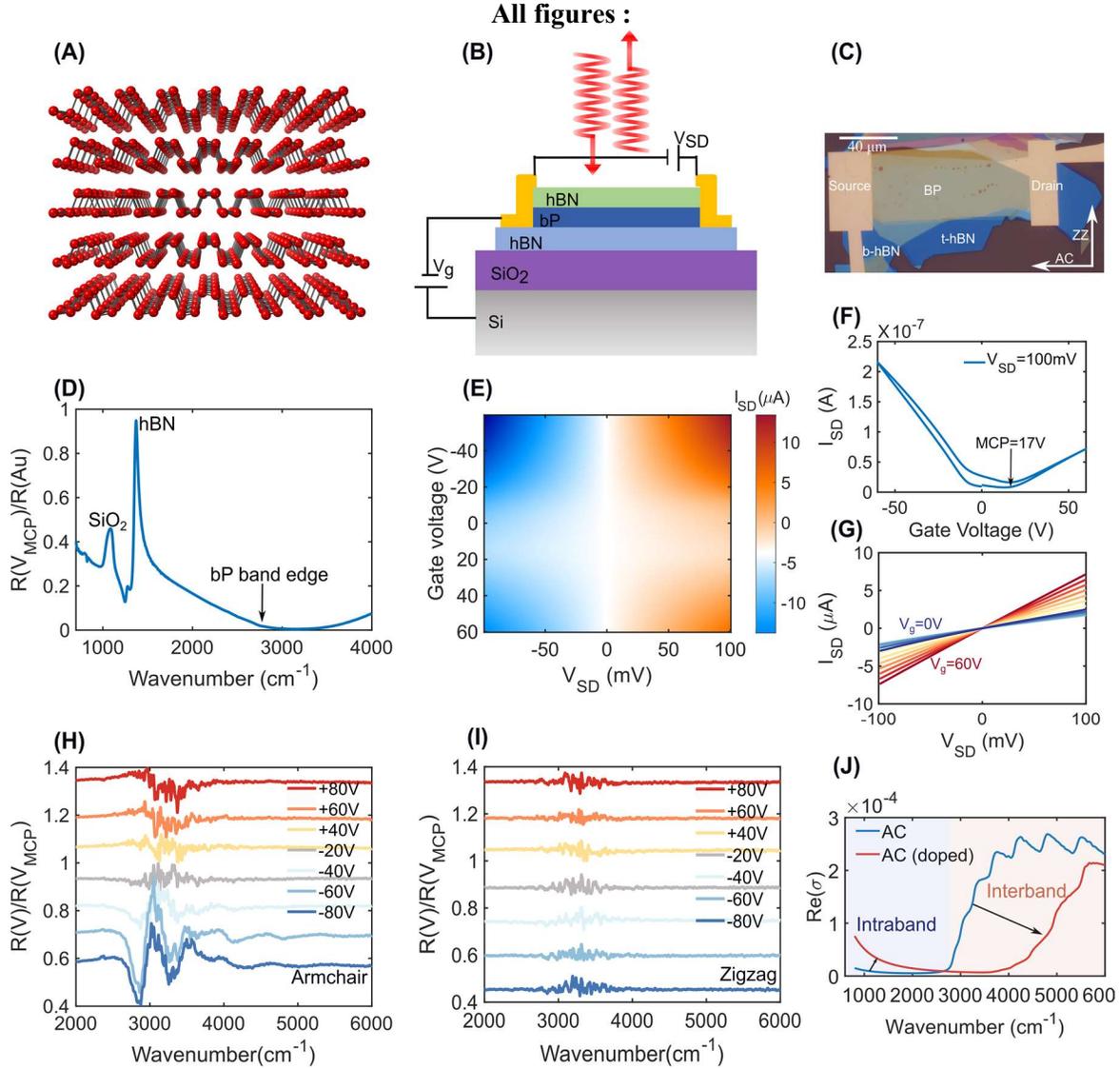

**Fig. 1. Device schematic and electro-optic characterization.** **(A)** Anisotropic puckered crystal structure of BP (P atoms are in sp$^3$ hybridization). **(B)** Device schematic and measurement scheme for hBN encapsulated BP devices. **(C)** Optical microscope image of the device discussed in the main text. **(D)** Normalized reflection spectrum from the BP device shown in **(C)**. **(E)** Color-map of source-drain current variation as a function of both gate voltage and source-drain bias. **(F)** Gate voltage modulated source-drain current at one representative source-drain voltage (100mV). **(G)** Variation of source-drain current with source-drain voltage showing linear conduction with systematic increase as gate voltage increases on the positive side, the slight dip is due to the fact that the MCP is not at 0V). **(H),(I)** Interband optical modulation along the AC and ZZ axis respectively showing the anisotropy in the electro-optic effects. **(J)** Schematic of changes in the AC axis optical conductivity (real part) upon doping.

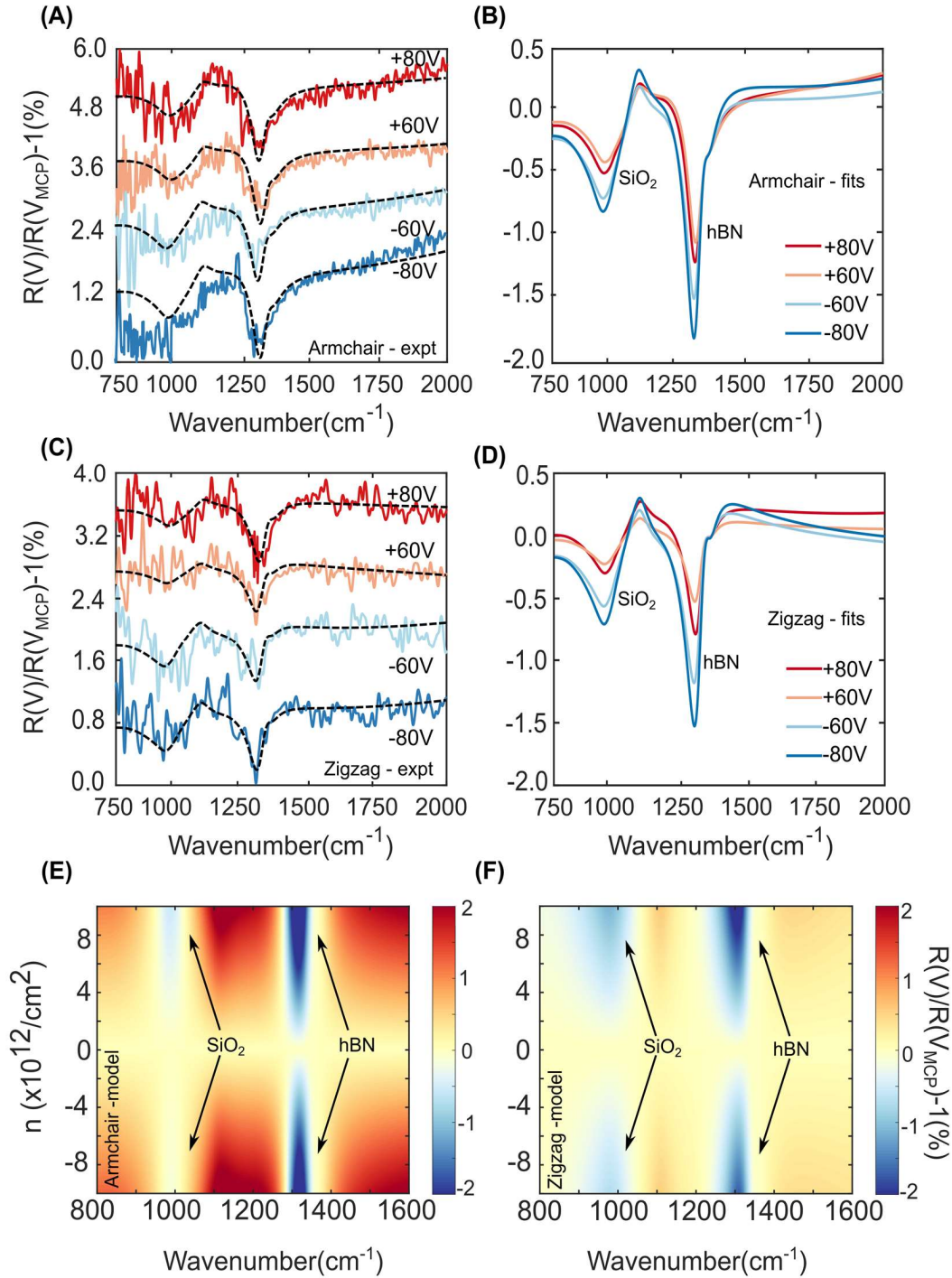

**Fig. 2. Intraband response dominated reflection modulation.** **(A),(B)** Measured (colored lines) and simulated/fit (black lines) intraband response mediated reflection modulation along the AC and ZZ axis. The fits have been performed between 750-2000 cm$^{-1}$ to eliminate any band-edge effect influence on the optical conductivity so that the Drude model suffices. **(C),(D)** Fits shown separately, without offset showing a narrowing and strengthening of the Fano-like response near the hBN and SiO$_2$ phonons with increasing charge density in BP. **(E),(F)** Modelled false color plot of modulation in reflection spectra (zoomed in between 800 and 1600 cm$^{-1}$) as a function of doping density for the AC and ZZ direction assuming the following parameters : BP m$_{eff}$=0.14m$_0$ (AC), 0.71m$_0$(ZZ), Si m$_{eff}$=0.26m$_0$ (electrons), 0.386m$_0$ (holes).

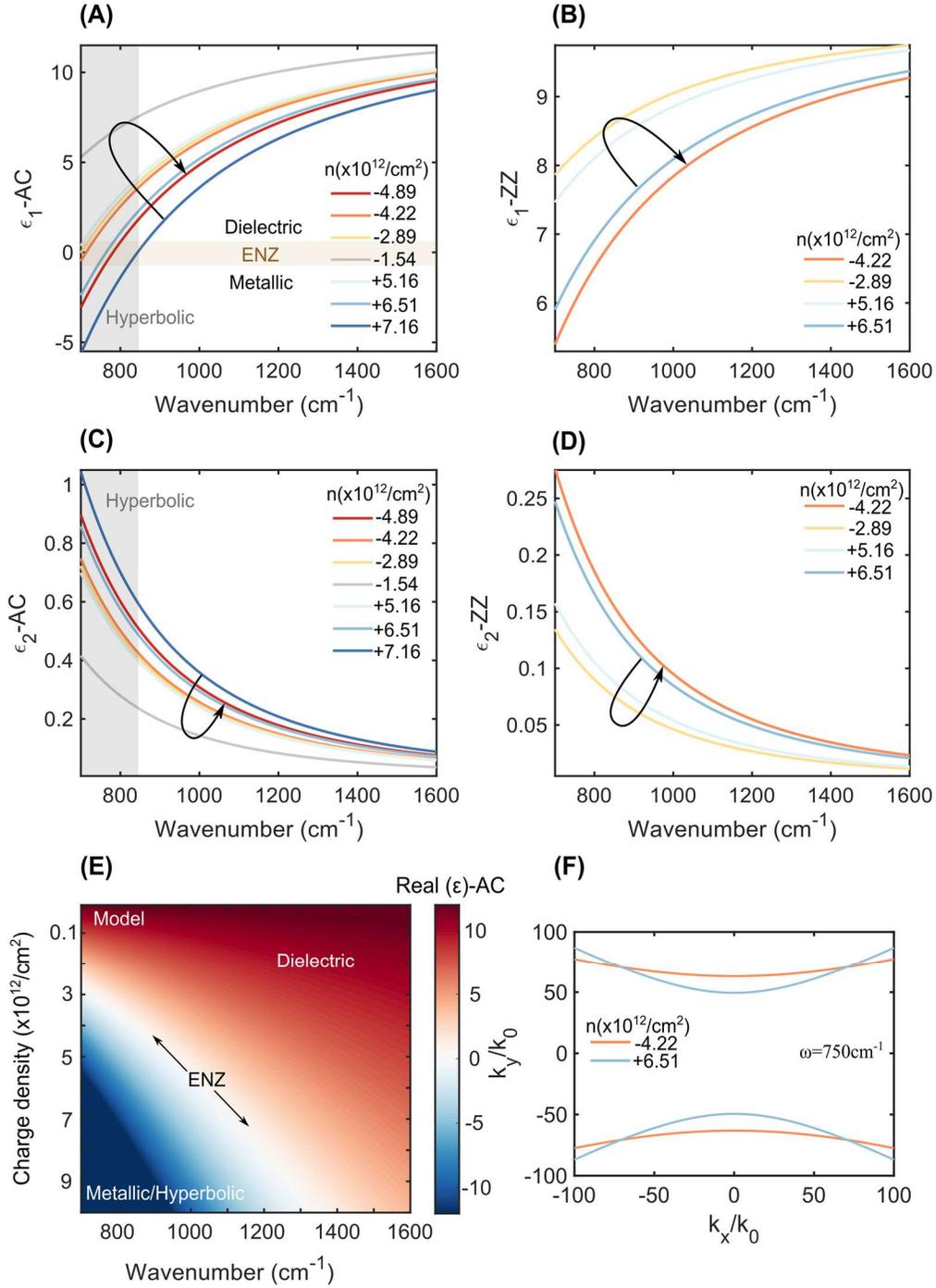

**Fig. 3. Modelled dielectric function and tunable hyperbolicity. (A),(C)** Extracted real and imaginary part of the dielectric function for BP 2DEG along the Armchair axis for different doping densities. The orange shaded region shows the ENZ behavior. The region where the real part of the permittivity along the AC axis goes negative while remaining positive for the ZZ direction is the hyperbolic region and extends to frequencies beyond our measurement window. **(B),(D)** The same for the Zigzag axis. **(E)** False color plot of the modelled real part of the dielectric permittivity along the AC direction assuming BP $m_{eff}=0.14m_0$ showing the tunability of ENZ. **(F)** Calculated isofrequency contours for in-plane plasmonic dispersion (TM polarized surface modes) showing the tunability of hyperbolicity.

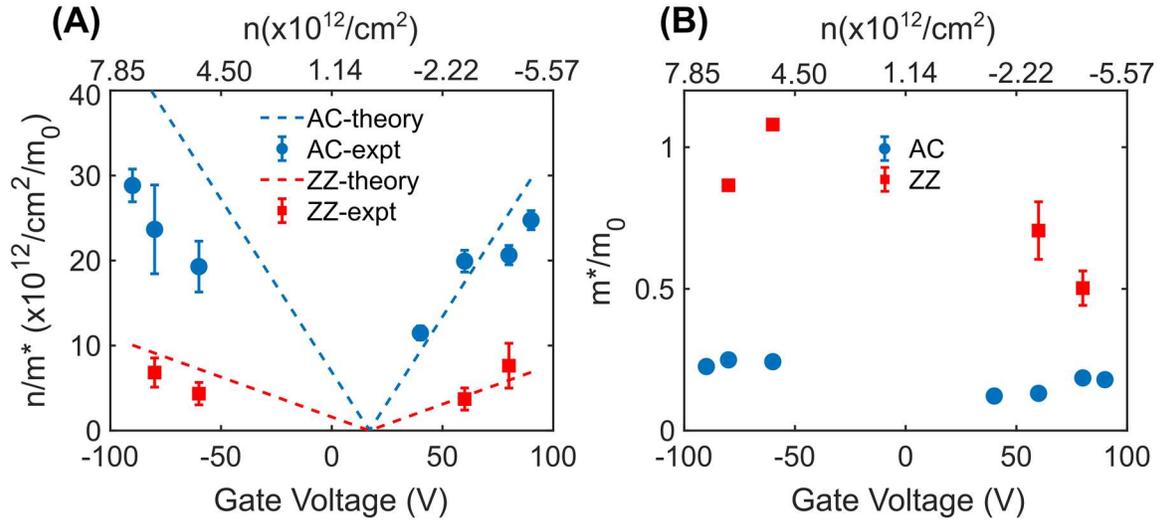

**Fig. 4. Extracted Drude weight and effective mass for BP. (A)** Drude weight evolution obtained from fitting reflection data for AC and ZZ axis, plotted with expected Drude weight. **(B)** Extracted effective mass from the Drude weight fits plotted versus voltage/charge density assuming a parallel plate capacitor model and 100% gating efficiency.